\documentclass[sigconf,submit]{acmart}
\AtBeginDocument{%
  \providecommand\BibTeX{{%
    \normalfont B\kern-0.5em{\scshape i\kern-0.25em b}\kern-0.8em\TeX}}}

\setcopyright{acmcopyright}
\copyrightyear{2022}
\acmYear{2022}
\acmDOI{XXXXXXX.XXXXXXX}

\acmConference[Conference acronym 'XX]{Make sure to enter the correct
  conference title from your rights confirmation emai}{June 03--05,
  2018}{Woodstock, NY}
\acmPrice{15.00}
\acmISBN{978-1-4503-XXXX-X/18/06}

\usepackage{tikz}
\usepackage{caption}
\usepackage{subcaption}
\usepackage{listings}
\usepackage[ruled,vlined]{algorithm2e}
\usepackage{multirow}
\usepackage{booktabs}
\usepackage{placeins}
\usepackage{float}
\begin{document}

\title{STELLA: Sparse Taint Analysis for Enclave Leakage Detection}
    

\author{Yang Chen$^\dagger$, Jianfeng Jiang$^\ddagger$,  Shoumeng Yan$^\ddagger$, Hui Xu$^\dagger$}
\authornote{Corresponding author.}
\affiliation{
  \institution{$^\dagger$School of Computer Science, Fudan University}
  \country{China}
}
\affiliation{
  \institution{$^\ddagger$Ant Group} 
  \country{China}
}

\begin{abstract}
Intel SGX (Software Guard Extension) is a promising TEE (trusted execution environment) technique that can protect programs running in user space from being maliciously accessed by the host operating system. Although it provides hardware access control and memory encryption, the actual effectiveness also depends on the quality of the software. In particular, improper implementation of a code snippet running inside the enclave may still leak private data due to the invalid use of pointers. This paper serves as a first attempt to study the privacy leakage issues of enclave code and proposes a novel static sparse taint analysis approach to detect them. We first summarize five common patterns of leakage code. Based on these patterns, our approach performs forward analysis to recognize all taint sinks and then employs a backward approach to detect leakages. Finally, we have conducted experiments with several open-source enclave programs and found 78 vulnerabilities previously unknown in 13 projects.

\end{abstract}


\maketitle

\section{Introduction}
Due to the advantages of cloud computing, migrating on-premise software to public cloud services has received more and more attention\cite{dong2022t}. However, privacy is still a major concern for many potential cloud users that cannot afford data leakage risks\cite{9095418}. TEE (trusted execution environment) is a promising solution to this problem provided by chip manufacturers, such as Intel Software Guard Extensions (SGX)~\cite{mckeen2013innovative,hoekstra2013using} and ARM TrustZone~\cite{holdings2009arm,ngabonziza2016trustzone}. With TEE, cloud users can deploy their security-critical applications in isolated enclaves and prevent them from being accessed by unauthorized parties. In recent years, we have witnessed the application of Intel SGX in many complex systems, including database \cite{priebe2018enclavedb,zhou2021veridb,wang2017cryptsqlite}, middleware \cite{brenner2016securekeeper,10.1145/3302424.3303951}, blockchain \cite{milutinovic2016proof,ayoade2018decentralized}, and network \cite{kim2018sgx,aublin2017talos}. 




Although Intel SGX provides hardware-level protection for the software, its practical effectiveness largely depends on how users code their programs. Any improper use of the SGX interface or bugs in the enclave code may lead to privacy leakage~\cite{randmets2021overview}. To our best knowledge, there is still neither systematic work on this problem nor available solutions to detect such bugs. Studying this issue is particularly important because verifying the effectiveness of TEE code is a critical step to the advancement of TEE solutions.

In this study, we first list five bug patterns that lead to privacy leaking in enclave programs (writing sensitive data to ECALL \texttt{out}, ECALL \texttt{user\_check}, OCALL \texttt{in}, OCALL return and NULL pointers). To the best of our knowledge, no research has summarized these patterns. A well-known technique for finding privacy leaks is taint analysis~\cite{arzt2014flowdroid,yang2012leakminer,celik2018sensitive}, and it is straightforward to apply taint analysis to find the bug patterns we define. However, enclave programs' privacy leak exhibits a variety of traits. For instance, taint sinks are data writes to particular pointers that are unknown and require solving. This is distinct from the typical one in that taint sinks are fixed and well-known APIs. Therefore, it is challenging to use taint analysis to tackle this issue.


To find these patterns in enclave programs, we provide an innovative custom taint analysis. We initially investigate the def-use relationship of the variables in enclave programs in order to construct a value flow graph (VFG). In order to identify the data writes to these pointers (i.e., identify taint sinks), we first perform forward analysis starting from the definition nodes of these pointers on the VFG. The written data is examined using backward analysis to determine its sensitivity. We employ VFG rather than control flow graph (CFG) as in conventional static analysis because VFG-based analysis (also known sparse analysis) has been proven in prior studies~\cite{10.1145/2338965.2336784,10.1145/2854038.2854043,shi2018pinpoint,10.1145/3453483.3454086,sui2016svf} to improve performance without sacrificing accuracy.


 To elaborate, the Enclave Definition Language (EDL)~\cite{intelSGXSDKDeveloperReference} files, which serve as interface descriptions, and the LLVM IR files generated by the enclave program are the inputs used by our approach. We parse EDL files to extract pointer parameters that could result in the leakage of private information and to create the VFG, we read the LLVM IR files. Then, starting with the definition nodes associated with the aforementioned pointer parameters, we traverse the VFG and taint the variables in accordance with the rules we established for taint propagation. We designate them as taint sinks if we discover writes to these taint variables in the enclave program. The VFG also provides the corresponding nodes for the data written in a taint sink. To obtain all the leaked variables, we begin from these nodes and move backward via the VFG. If any of them are private, it suggests a possible privacy breach. Our technique provides manual insensitive annotation at the source code level and insensitive variables will be automatically omitted in the backward analysis to reduce false positives because not all variables in the enclave program are sensitive. Furthermore, the decrypted data, the data to be encrypted, and the key for encryption or decryption will be automatically tagged as high-risk sensitive data. In the leak report, the high-risk data will be shown with a greater priority if it has been compromised.

We have implemented a prototype for the \textbf{S}parse \textbf{T}aint analysis for \textbf{E}nc\textbf{L}ave privacy \textbf{L}e\textbf{A}kage detection, namely STELLA. In addition to employing sparse analysis to increase speed, we have adopted a preference for retrieving the call graph (CG) rather than the call flow graph (CFG) for solving variables on VFG that could reveal secrets because CFG has a significantly higher number of nodes than CG. We used STELLA to analyze 13 open-source SGX-based applications on GitHub to test the viability of our approach, and we discovered 78 privacy disclosure bugs there that had never before been reported.

In short, this article contains several major research contributions as follows.
\begin{itemize}
\item Privacy leaks caused by coding errors in enclaves undermine enclave confidentiality guarantees, making it easier for attackers to obtain secrets in enclaves. To the best of our knowledge, our work is the first comprehensive one on this issue.
\item We first define patterns of privacy-leaking coding mistakes that enclave developers are prone to make and then propose a novel sparse taint approach to efficiently analyze the vulnerabilities in enclave code based on these patterns. As far as we know, this is the first available approach on the issue.
\item We have implemented a prototype tool, STELLA, and released it as open-source on GitHub. Our experimental results show that STELLA can effectively discover privacy-leaking vulnerabilities in enclave programs.
\end{itemize}

The rest of our paper is organized as follows. We introduce the Intel SGX background in section 2. Section 3 defines the problem of privacy leakage on enclave programs and discusses the challenges of detecting such bugs. Section 4 then elaborates on our sparse taint analysis approach. Section 5 evaluates the performance of our approach. We review related work in Section 6. Section 7 concludes our paper.




\section{Preliminary}
In this section, we first introduce background about SGX and Intel SGX SDK, then illustrate the privacy leakage issue when developing SGX programs.

\subsection{SGX Background}

Intel SGX is a set of CPU instructions that enable applications to create hardware-based protected areas, namely enclaves. Enclaves are used to protect private data from modification and closure~\cite{intel-sgx-sdk}. Private data within an enclave can only be accessed by the code within the same enclave, and cannot be read or written by programs outside the enclave even such programs run at privilege levels.

\begin{figure}
\centering
\includegraphics[width=0.47\textwidth]{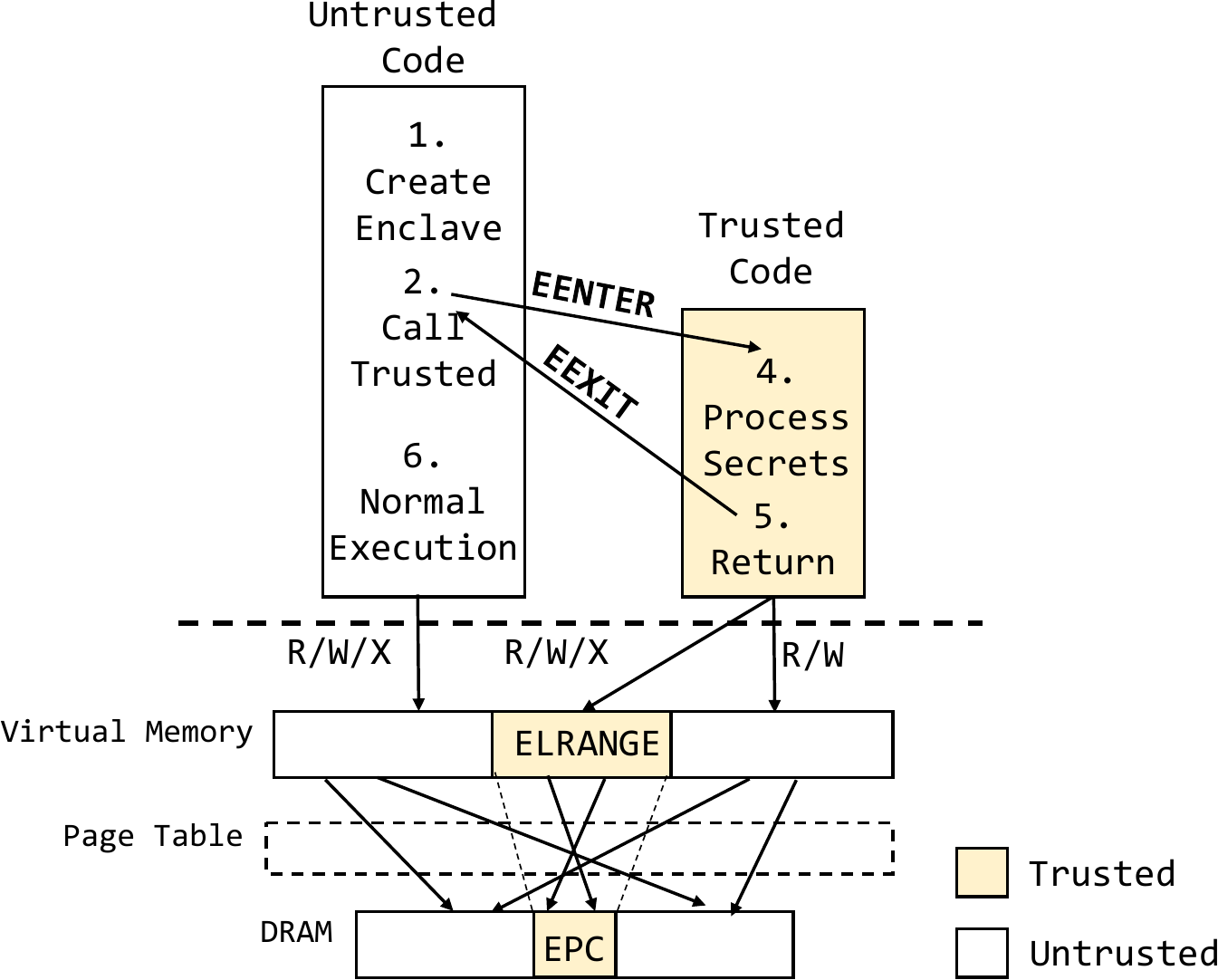}
\caption{The underlying mechanism of SGX.}
\label{fig:SAEFAMP}
\end{figure}
A big change in introducing enclaves in a normal program is that we need to separate the application into the untrusted part (outside enclave) and trusted part (inside enclave) and all access to secrets needs to be done in the enclave. As shown in Figure \ref{fig:SAEFAMP}, untrusted code can create one or more enclaves. When the secrets need to be processed, the untrusted code needs to call the trusted function like a normal function, and then the CPU is switched to the enclave mode via the EENTER instruction, and the control is transferred to the corresponding trust function. The trusted function in the enclave can directly access these secrets. After processing these secrets, the CPU switches out of enclave mode via EEXIT instruction, and the enclave transfers control to untrusted code and continues normal execution.

\subsection{Intel SGX SDK}
Intel SGX SDK \cite{intelSGXSDK} is a set of toolkits to develop enclave programs. In practice, developers tend to write programs with high-level languages, instead of directly using low-level instructions like EENTER and EEXIT. To this end, SGX SDK provides high-level abstractions on low-level instructions and enables developers to develop enclave programs with C/C++. Currently, enclave programs are mostly developed with Intel SGX SDK.

\begin{figure}[t]
\centering
\includegraphics[]{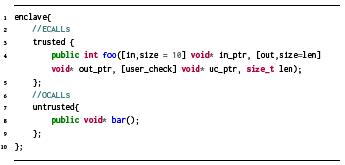}
\caption{An example EDL file.}
\label{fig:example_edl}
\end{figure}

Intel SGX SDK introduces ECALLs (enclave calls, the calls to enclave functions) and OCALLs (outside calls, the calls to outside-enclave functions) on top of the EENTER and EEXIT instructions. An ECALL calls the EENTER instruction first, then executes the trusted function, and finally calls the EEXIT instruction. OCALL is vice versa. ECALLs and OCALLs are the interfaces between host and enclave, and their prototypes are defined in EDL files. As shown in Figure \ref{fig:example_edl}, we declare an ECALL \texttt{foo} and an OCALL \texttt{bar} in the EDL. The trusted ECALL \texttt{foo} is defined in the enclave. The untrusted OCALL is defined in the host. 

\begin{figure}
\centering
\includegraphics[width=0.45\textwidth]{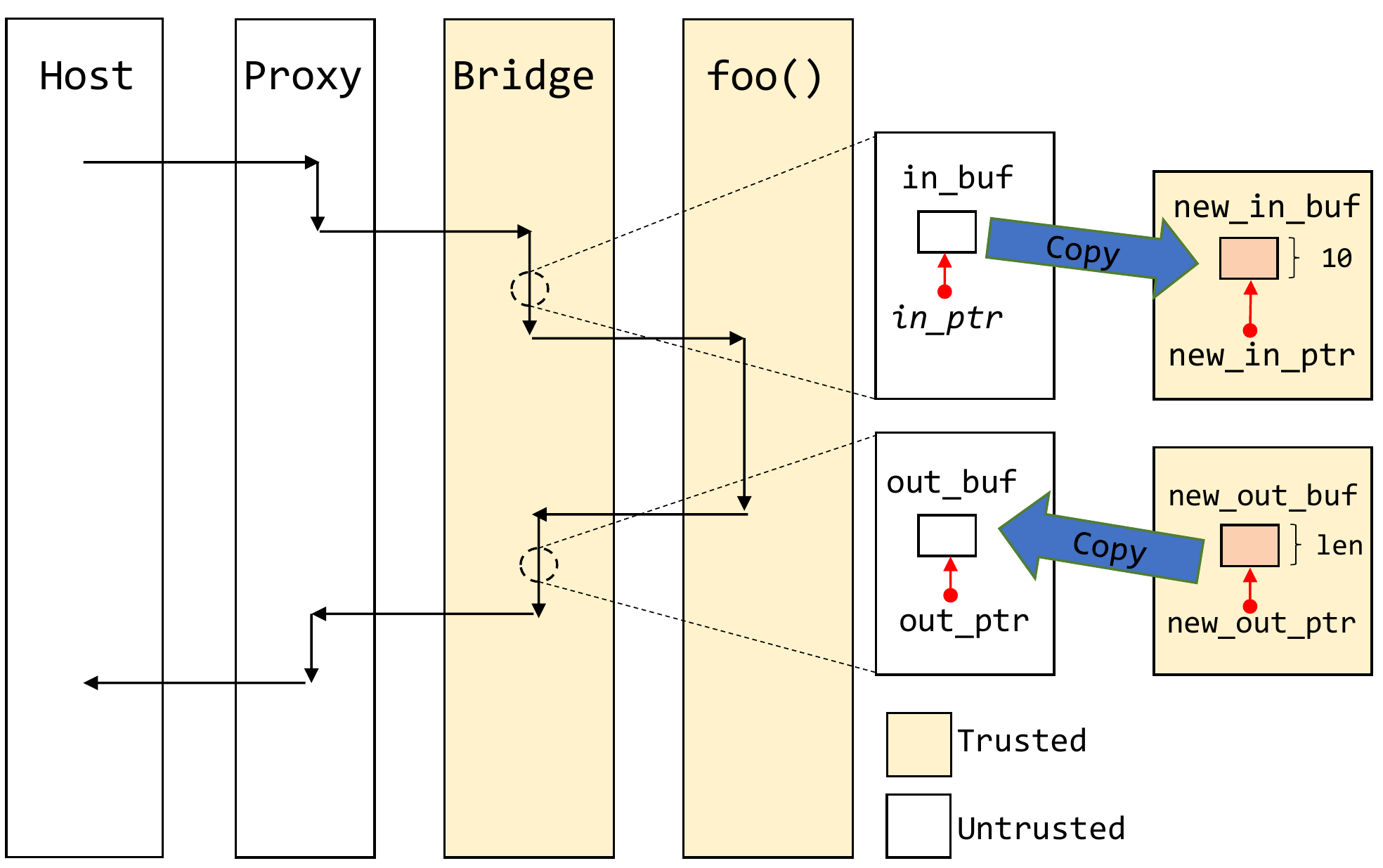}
\caption{Proxy and bridge functions for the ECALL foo.}
\label{fig:Proxy_and_bridge_functions_for_an_ECALL}
\end{figure}
Intel SGX SDX automatically generates proxy and bridge functions for ECALLs and OCALLs when compiling enclave programs. Take the ECALL \texttt{foo()} in Figure \ref{fig:example_edl} as an example, we call its definition in enclave real \texttt{foo()}. Intel SGX SDK will generate a pair of functions, an untrusted proxy and a trusted bridge function. When executing this ECALL as Figure \ref{fig:Proxy_and_bridge_functions_for_an_ECALL} shows, The host code first calls the untrusted proxy. Then the untrusted proxy will call the trusted bridge. Finally, the trusted bridge calls the real foo(). The state and return value of the ECALL are propagated back from the opposite direction.

There are two categories of pointer parameters for functions defined in EDL files, namely directed pointers, and raw pointers. 

Pointer parameters in ECALLs and OCALLs can be declared with direction (\texttt{in}, \texttt{out}) and size (\texttt{size\_t}). The proxy and bridge functions will copy the content of pointers based on the direction and size. Take \texttt{foo} in Figure~\ref{fig:example_edl} as an example, \texttt{in\_ptr} has the \texttt{in} attribute, and its size attribute is 10. The trusted bridge function will allocate a 10-bytes buffer inside the enclave and copies the first 10 bytes of \texttt{in\_ptr} to the buffer. Similarly, if the direction is \texttt{out}, the trusted proxy will copy data from the enclave to the the untrusted world.

Pointers can also be annotated as \texttt{user\_check} (\textit{e.g.} \texttt{uc\_ptr} in \texttt{foo} in Figure~\ref{fig:example_edl}), which means it is a raw pointer. The bridge or proxy will not copy the buffer but directly pass the address. Note that the pointer returned by an OCALL, like \texttt{bar} in Figure \ref{fig:example_edl}, is also a raw pointer, and developers need to do proper checks before using it. In addition, if an \texttt{in} pointer is a pointer to struct, since the buffer copy is a shallow copy, the pointer fields in the struct are all \texttt{user\_check} pointers.

In general, the pointer direction attributes in EDL determine whether buffer copying or raw pointer transfer occurs between host and enclave. If developers are not careful with these pointers, sensitive data can be accidentally leaked into the untrusted world. In the following subsection, we will demonstrate this problem with a real-world example.

\subsection{Issues of Developing with SGX SDK}

Although SGX prevents host code from directly accessing data inside the enclave, it is still possible to leak data outside the enclave through the pointer parameter of ECALL or OCALL. For example, writing sensitive data to the \texttt{in} pointers in OCALLs may cause privacy leakage.

\begin{figure}[t]
\centering
\includegraphics[]{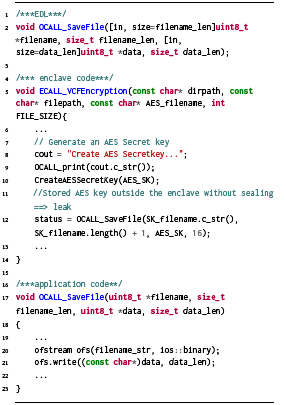}
\caption{Motivating example: AES key leakage in BiORAM-SGX.}
\label{fig:Motivating_Example}
\end{figure}

Such privacy leakage problems are detected in existing enclave programs. BiORAM-SGX~\cite{BiORAM-SGX} is a personal genetic data statistical analysis system using SGX. The system claims that sensitive personal genetic data will not be leaked to the server. But we found the problem of AES key leakage in this system. Figure \ref{fig:Motivating_Example} demonstrates the buggy code snippet. \texttt{data} of \texttt{OCALL\_SaveFile} is an OCALL \texttt{in} pointer (line 2), which can pass information from the enclave to the outside of the enclave. The developers must be careful with this pointer and cannot pass confidential information to \texttt{data}. However, \texttt{ECALL\_VCFEncryption} passes the AES key to \texttt{data} in plaintext without sealing, calling \texttt{OCALL\_SaveFile} to save the AES key out of the enclave (line 12). An attacker can easily open the unencrypted file and obtain the AES key, which poses a great threat to user data security. 

Note that this vulnerability does not contradict SGX's guarantee against data leakage. SGX SDK requires developers to perform checks when writing sensitive data to pointers. If the developers carelessly fail to check, SGX cannot guarantee that the data inside the enclave will not be leaked.

\section{Problem and Challenges}
For enclave programs, developers may mistakenly copy sensitive data out of the enclave, which will cause privacy leakage. Note that the privacy leakage problem is serious for enclave programs because this problem breaks SGX's guarantee of data security. SGX aims to protect private data from access from the host and not allow the leaking of private data to the untrusted world.

Next, we first summarize common code patterns that may cause the problem, then illustrate the challenges to adapt taint analysis to enclave programs.

\subsection{Typical Patterns}
\begin{figure}
\centering
\includegraphics[width=0.4\textwidth]{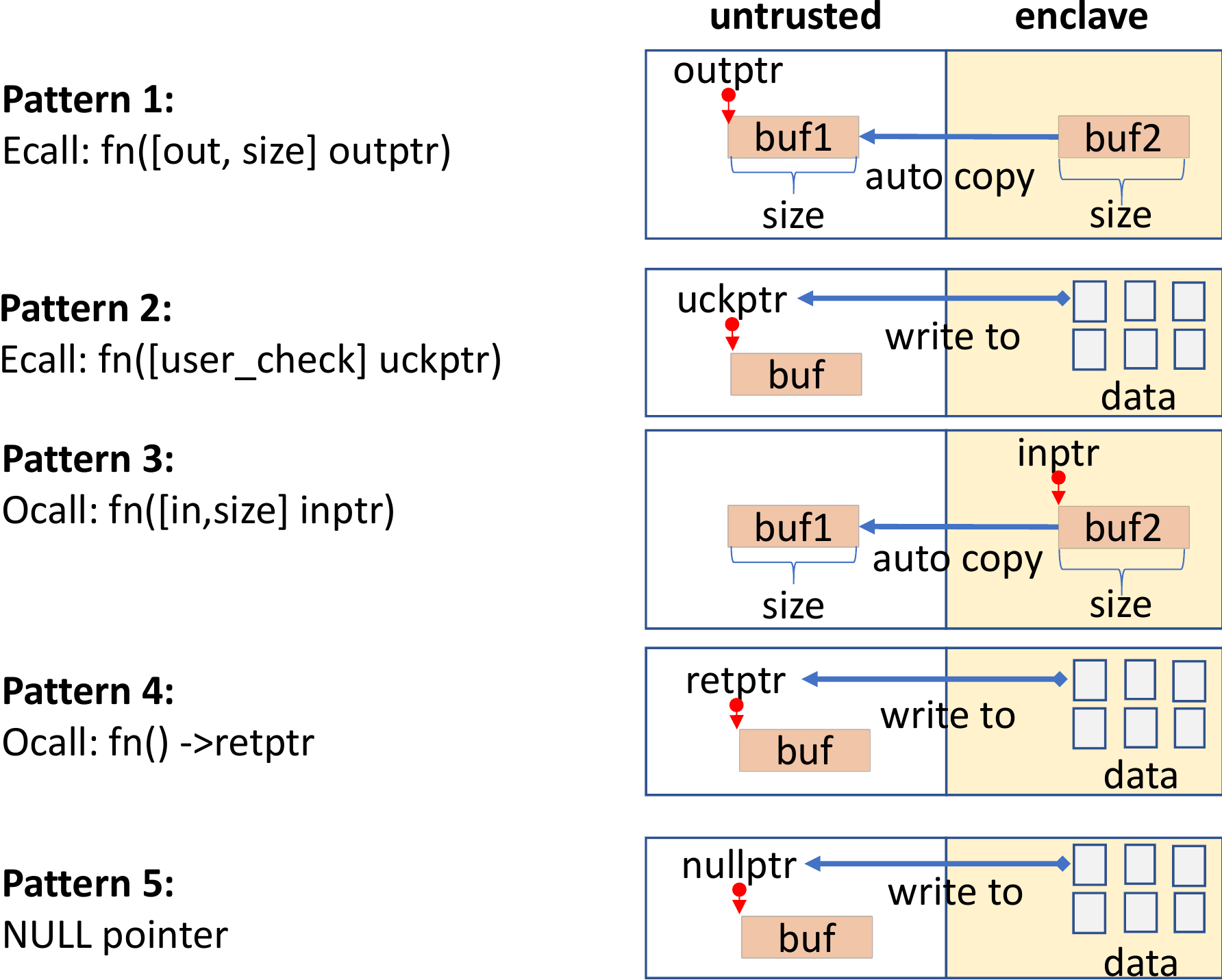}
\caption{Writing sensitive data to these five pointers in the enclave program will cause privacy leakage. The (ECALL \texttt{out} / OCALL \texttt{in}) pointer points to the variable in the enclave, but the written data will be copied outside the enclave by the SGX SDK. The (ECALL \texttt{user\_check} / OCALL return / NULL) pointer points directly outside the enclave.}
\label{fig:five_pointers}
\end{figure}

For enclave programs, most privacy leakages are caused by writing sensitive data to specific pointers. As shown in Figure \ref{fig:five_pointers} , by comprehensively studying the coding specification of Intel SGX SDK, we summarize 5 pointer usage patterns that can leak privacy in enclave codes. To the best of our knowledge, no one has yet summarized these patterns.

\textbf{Write Private Data to ECALL \texttt{out} Pointers (P1)}. As shown in Figure \ref{fig:Proxy_and_bridge_functions_for_an_ECALL}, When ECALL \textit{foo} returns, the data written to the ECALL out pointer is copied into untrusted memory by the bridge function. Therefore, when a developer mistakenly writes private data to an ECALL \texttt{out} pointer, the private data is leaked.

\textbf{Write Private Data to ECALL \texttt{user\_check} Pointers (P2)}. To achieve better performance, developers may use \texttt{user\_check} pointers in ECALL to avoid checking pointers and copying buffer by SGX SDK. Instead, developers need to check the \texttt{user\_check} pointer by themselves. But they may carelessly not check \texttt{user\_check} pointers and write sensitive data to them. If the pointers point to untrusted memory, private data may leak. Note that we cannot trust the ECALL input according to the Intel SGX threat model, so even if enclave developers assume that these pointers point inside the enclave, they may also be tampered with by an attacker to point to untrusted memory. Note that the pointer field in the \texttt{in} structure pointer is \texttt{user\_check}, so this pattern also includes this case.

\textbf{Write Private Data to OCALL \texttt{in} Pointers (P3)}. Figure \ref{fig:Proxy_and_bridge_functions_for_an_ECALL} shows how Intel SGX SDK processes the \texttt{in} pointer in ECALL. The bridge function copies the untrusted memory data to the enclave. For the \texttt{in} pointer of OCALL, the processing is similar but in the opposite direction. The Intel SGX SDK will copy the enclave memory data to the untrusted memory, so developers must be careful, if the OCALL \texttt{in} pointer points to sensitive data, it will lead to privacy leakage.

\textbf{Write Private Data to OCALL Return Pointers (P4)}. If the return pointer of OCALL points to untrusted memory, and the developer mistakenly writes sensitive data to this pointer, it will lead to privacy leakage.

\textbf{Write Private Data to NULL Pointers (P5)}. For general applications, writing data to address 0 (NULL) will be aborted by the OS. However, for SGX applications, the OS is untrusted according to the Intel SGX threat model. Writing sensitive data to NULL pointers in the enclave can lead to privacy leakage.

Taint analysis is a traditional method to detect such patterns. However, since enclave programs differ from traditional programs in many ways, we have to adapt the taint analysis to enclave programs.

\subsection{Challenges}
The traditional method to detect privacy leakage is via taint analysis, \textit{e.g.,} for android or IoT programs. However, traditional taint analysis cannot be directly ported to detecting privacy leakage on enclave programs. A traditional taint analysis framework includes \textbf{taint sources}, \textbf{propagation rules}, and \textbf{taint sinks}. Enclave programs differ from android or IoT programs in the following three aspects:

\textbf{Taint sources} are variables storing sensitive data. In general, android programs only have a few taint sources such as those variables storing passwords and phone numbers so manually annotating such variables is easy. For enclave programs, considering that SGX protects all data inside the enclave, we should treat all of them except a little public data as taint sources. Hence, there are two main challenges to our task. First, the taint sources will be relatively large. Manually marking taint sources is labor-intensive. Second, if all data inside the enclave is marked as sensitive data, we will have many false positives. we should distinguish between sensitive data and public data to avoid false positives.

\textbf{Propagation rules} define how tainted variables are propagated to other untainted variables through CFG. For example, if we convert a tainted variable into its hexadecimal format and store it in another variable, then the latter variable should also be tainted. For enclave programs, encryption functions should be considered during designing propagation rules, and only the leakage of unencrypted data should be viewed as private data leakage. For example, If the parameters of OCALLs are encrypted by some functions \textit{e.g.,} \texttt{sgx\_rijndael128GCM\_encrypt}, it should not be considered a privacy leakage.

\textbf{Taint sinks} are usually APIs that may leak information. For android programs, taint sinks are fixed, like APIs sending SMS messages or remote requests. However, for enclave programs, taint sinks are related to some specific pointers, rather than fixed functions. We have to find pointers that can leak data out of the enclave. Codes that dereference or write to these pointers are the taint sinks we need.

Taint analysis tries to find paths from taint sources to taint sinks. If such a path exists, private data may be leaked.

\section{Sparse Taint Analysis Approach}


\begin{figure}
\centering
\includegraphics[width=0.478\textwidth]{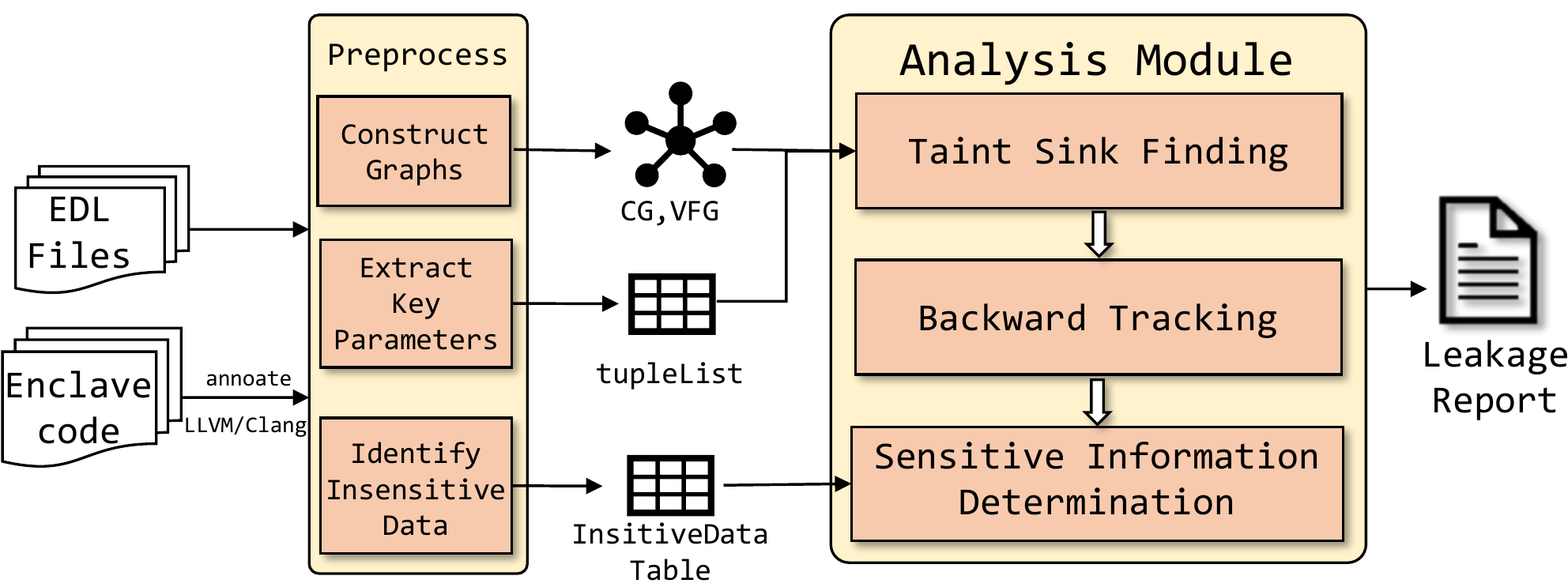}
\caption{Overall framework of STELLA.}
\label{fig:Overall_framework_of_STELLA}
\end{figure}


In general, we use static analysis on SGX programs to detect privacy leakage vulnerabilities. The whole STELLA framework is shown in Figure \ref{fig:Overall_framework_of_STELLA}. STELLA receives EDL files and annotated enclave code as input and generates a leakage report. Preprocessing and analysis module comprise the STELLA pipeline.

\subsection{Preprocess}
As shown in Figure \ref{fig:Overall_framework_of_STELLA}, STELLA first preprocesses input files, including constructing graphs, extracting key function parameters, and identifying insensitive data. The module accepts compiled bitcode and EDL files as input and outputs graphs and tables required for further analysis.
\subsubsection{Construct Graphs}

We construct several graphs for analysis, including CG and VFG. STELLA bases graph construction on SVF~\cite{sui2016svf}, a static value flow analysis framework. VFG plays a key role in our analysis. STELLA performs an inter-procedural pointer analysis to obtain the points-to information after reading the enclave bitcode files. With the points-to information, STELLA constructs a memory SSA form and obtains the def-use chains and value flows to construct the VFG.

\begin{figure}[t]
\centering
\begin{subfigure}{0.45\textwidth}
\includegraphics[]{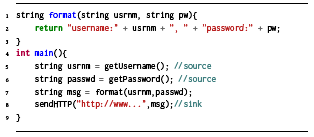}
\caption{Program Code}
\label{fig:Program_Code}
\end{subfigure}
\hfill
\begin{subfigure}{0.22\textwidth}
\includegraphics[width=1\textwidth]{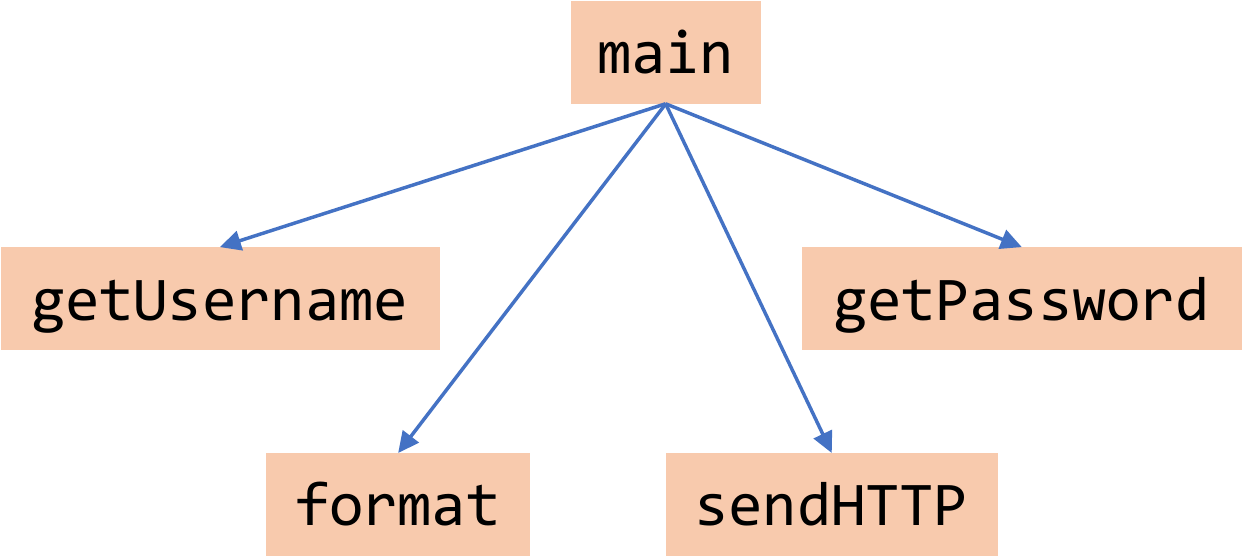}
\caption{Call Graph (CG)}
\label{fig:Call_Graph}
\end{subfigure}
\hfill
\begin{subfigure}{0.22\textwidth}
\centering
\includegraphics[width=1\textwidth]{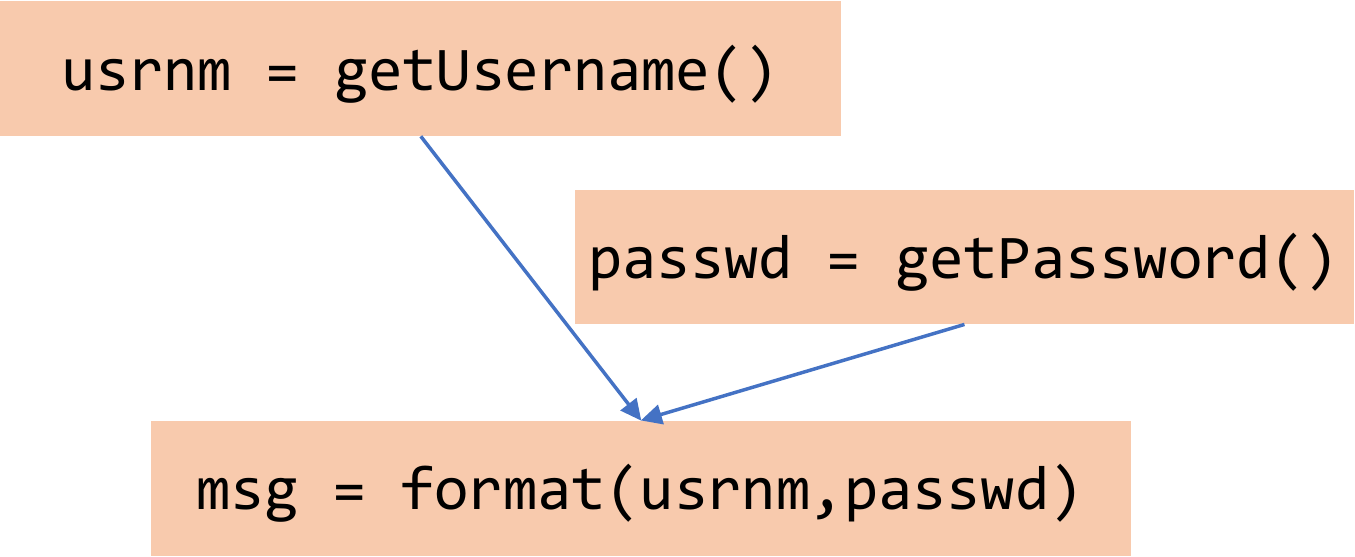}
\caption{Value Flow Graph (VFG)}
\label{fig:Value_Flow_Graph}
\end{subfigure}
\caption{Simple taint and source-sink example code and its graphs}
\label{fig:Simplest_taint_and_source_sink_example_code_and_its_graphs}
\end{figure}

Figure~\ref{fig:Simplest_taint_and_source_sink_example_code_and_its_graphs} demonstrates CG and VFG of an example program. Figure \ref{fig:Program_Code} is a simple taint source-sink example program. The program fetches the user name and password, assembles a message, and finally sends it to the outside via HTTP, causing sensitive information leakage. Call graph (CG) describes the calling relationship between functions. As shown in Figure \ref{fig:Call_Graph}, the \texttt{main} function calls \texttt{getUsername}, \texttt{getPassword}, \texttt{format} and \texttt{sendHTTP}. Value Flow Graph (VFG) demonstrates the data flow of variables. As shown in Figure \ref{fig:Value_Flow_Graph}, The values of variables \texttt{usrnm} and \texttt{passwd} both flow to variable \texttt{msg}. Our approach is based on CG and VFG. In general, we begin by matching CG nodes with the function name. The taint sink nodes on the VFG are then discovered using these nodes. Finally, leakage of personal information is discovered by resolving the graph reachability problem between sinks and sources. 

\subsubsection{Extract Key Parameters}
We extract the function parameters in ECALLs and OCALLs in EDL files that may potentially leak privacy. For each parameter, we combine its function name, the position index, and the leak pattern into a tuple. For example, for an ECALL, \texttt{ecall\_func([user\_check] void* uc\_ptr, [out,size = size] void* out\_ptr, int size)}, \texttt{uc\_ptr} and \texttt{out\_ptr} are two parameters that may leak privacy, so we will get two tuples, \texttt{(ecall\_func,0,P2)} and \texttt{(ecall\_func,1,P1)}. After scanning all EDL files, we will get a list of tuples, namely \texttt{tupleList}. It is used for subsequent taint analysis to find sinks. 
 
\subsubsection{Identify insensitive Data}
\begin{figure}[t]
\centering
\includegraphics[]{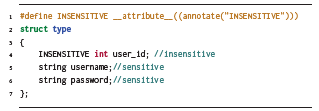}
\caption{Annotate the insensitive variable user\_id with INSENSITIVE. Other variables are sensitive by default.}
\label{fig:annotated_code}
\end{figure}

We manually annotate some insensitive variables in enclave programs. Since enclaves are used to protect private data, we assume that most data processed inside enclaves is sensitive. However, some variables storing insensitive information are used to interact with the untrusted world, like \texttt{user\_id} in Figure \ref{fig:annotated_code}. We will annotate this variable with the \texttt{INSENSITIVE} prefix. We assume all variables without the annotation store sensitive data. \texttt{llvm.var.annotation} \cite{LLVMLanguageReferenceManual} is a LLVM intrinsic function. We can use this function to annotate local variables with arbitrary strings. For each variable annotated with \texttt{INSENSITIVE}, after its memory allocation statement (such as \texttt{alloca}), the \texttt{llvm.var.\\annotation} function is called once with the variable as the first argument. So we traverse the call graph to find all the code that calls \texttt{llvm.var.annotation}, and check whether the function has the argument "INSENSITIVE". Finally, we can collect all annotated variables into the \texttt{InsensitiveDataTable}.


\subsection{Analysis Module}
The analysis module conducts analysis to identify potential privacy leakage paths. It comprises of taint sinks finding, backward tracking, and sensitive information determination.
\subsubsection{Taint Sink Finding}
We divide the taint sinks in enclave code into three categories, \textit{Pointers from Outside}, \textit{Pointers Declared Inside} and \textit{Explicit Sinks}. We design different search strategies for each category.
\begin{itemize}
\item\textbf{Pointers from Outside}. From an enclave developer's perspective, ECALL \texttt{out} (P1) , ECALL \texttt{user\_check} (P2), and OCALL return pointers (P4) all come directly from outside the enclave. The strategies for detecting their associated taint sinks are also similar. Note that the enclave code may derive new pointers from these pointers. For example, if an ECALL passes in a \texttt{user\_check} secondary pointer (e.g. \texttt{[user\_check]void** p1}), then this pointer is dereferenced to get a new pointer \texttt{void* p2 = *p1}. \texttt{p2} may also point outside the enclave. Writing sensitive data to \texttt{p2} may also lead to privacy leakage. Therefore, to make our analysis as sound as possible, we search not only these pointers but also the pointers derived from them via a taint-style algorithm. The instruction to write data to the pointer in the search result is the taint sink.

\item\textbf{Pointers Declared Inside.} Pointers defined inside the enclave can also be corrupted, such as null pointers or wild pointers. These pointers may point outside the enclave. We define writing instructions to these pointers as taint sinks. We focus on the privacy disclosure vulnerability caused by writing sensitive data to a NULL pointer in an enclave (\textit{i.e.}, P5).

\item\textbf{Explicit Sinks.} Calling OCALL functions that accept primitive types or \texttt{in} pointers as parameters can leak private data (\textit{i.e.}, P3). We call this type of taint sinks \textit{Explicit Sinks} because they can be located directly by matching function names in the call graph. 

\end{itemize}

\begin{algorithm}[t]
\scriptsize
	\caption{Taint sinks finding}
	\label{algo:findSinks}
	\LinesNumbered
	\SetKwFunction{FindSinks}{FindSinks}
	\SetKwFunction{PtrTaint}{PtrTaint}
	\SetKwProg{Fn}{Function}{}{end}	
    \SetKwData{EDLs}{EDLs}
    \SetKwData{prog}{prog}
    \SetKwData{sinks}{sinks}
    \SetKwData{tupleList}{tupleList}
    \SetKwData{taintedPtrSet}{taintedPtrSet}
    \SetKwComment{Comment}{/* }{ */}
    
    \BlankLine
	\BlankLine
    \KwIn{\EDLs: EDL files, \prog: the enclave program to be analyzed}
    \KwResult{\sinks: taint sinks}
    
    \BlankLine
    \BlankLine
     \Fn{\FindSinks{\EDLs, \prog}}{
        $tupleList \gets \texttt{ExtractKeyParameters}(EDLs)$\\
        $cg,vfg \gets \texttt{ConstructGraphs}(prog)$\\
        $sinks \gets \{\}$\\
        \For{$tuple \in tupleList$}{
            /* match the cg node by funcname. */\\
            $node \gets \texttt{GetNode}(tuple.funcname,cg)$\\
                    /* Pointers from Outside (ECALL out / user\_check / OCALL return)*/\\
                    \If{$tuple.leakType \in \{P1,P2,P4\}$}{
                          $ptr \gets \texttt{GetPtr}(node,tuple.index)$\\
                          /* ptrTaint is shown in Algorithm \ref{algo:ptrTaint} */\\
                          $sinks \gets sinks \cup \texttt{PtrTaint}(ptr)$\\
                    }
                    /* Explicit Sinks (OCALL in pointers)*/\\
                    \ElseIf{$tuple.leakType \in \{P3\}$}{
                        $ptr \gets \texttt{GetPtr}(node,tuple.index)$\\
                        $defNode \gets \texttt{GetDefNode}(ptr,vfg)$\\
                        $sinks \gets sinks \cup \{defNode\}$\\
                    }
                }
        /* for NULL pointers (P5). */\\
        /* malloc without check */\\
        $node \gets \texttt{GetNode}("malloc",cg)$\\
        $ptr \gets \texttt{GetPtr}(node)$\\
        $cmpInsts \gets \texttt{GetCmpInsts}(ptr)$\\
        /*if no cmp instruction : no NULL check*/\\
        \If{$cmpInsts == \emptyset$}{
            $sinks \gets sinks \cup \texttt{PtrTaint}(ptr)$\\
        }
        \KwRet{$sinks$}
     }
\end{algorithm}

Algorithm \ref{algo:findSinks} demonstrates our taint sinks finding algorithm. We iterate over the entries of the \texttt{tupleList} and use the function name in the entry to match the node on the call graph. For \textit{Pointers from Outside}, we use the node and the parameter index to get the pointer to analyze and then call \texttt{PtrTaint} (Algorithm \ref{algo:ptrTaint}) to search for the taint sinks associated with this pointer. For \textit{Explicit Sinks}, it is relatively simple. After we locate the pointer parameter, its definition node on the VFG is the taint sink. For \textit{Pointers Declared Inside}, we focus on NULL pointers due to the enclave not validating the result of \texttt{malloc}. After we get the \texttt{malloc} pointers in the program, check whether they have corresponding comparison instructions. If not, it may be NULL, and then we call \texttt{PtrTaint} to search for taint sinks.

\subsubsection{Pointer Tainting}
We use a taint-style algorithm to search leak-causing pointers and derivative pointers and their related taint sinks. We design the following 6 propagation rules, which are represented by 6 formulas. The upper part of each formula represents the current situation or conditions, and the lower part of the formula represents the corresponding operation that needs to be performed. \texttt{T(v)} represents marking variable \texttt{v} as tainted, and \texttt{U(v)} represents removing the taint flag of the variable \texttt{v}.

\begin{equation}
\frac{T(addr),v = load(addr)}{T(v)}
\label{formula:load}
\end{equation}

\begin{equation}
\frac{T(v),store(v,addr)}{T(addr)}
\label{formula:store_op}
\end{equation}

\begin{equation}
\frac{T(x),y = op(x)}{T(y)}
\label{formula:unary_op}
\end{equation}

\begin{equation}
\frac{T(x),z = op(x,y)}{T(z)}
\label{formula:binary_op}
\end{equation}

\begin{equation}
\frac{T(base),addr = gep(base)}{T(addr)}
\label{formula:gep_op}
\end{equation}

\begin{equation}
\frac{T(x),y = bitcast(x)}{T(y)}
\label{formula:bitcast_op}
\end{equation}

In the formula (\ref{formula:load}), the LLVM load instruction refers to reading a value from memory. When \texttt{addr} is a tainted variable, then the value \texttt{v} read from \texttt{addr} is also marked as a tainted variable.

In the formula (\ref{formula:store_op}), the store instruction is used to write memory. The store instruction has two arguments: a value \texttt{v} to store and an address \texttt{addr} which indicates the write location. If \texttt{v} is a tainted variable, then after executing the store instruction, \texttt{addr} should also be marked as a tainted variable.

In the formula (\ref{formula:unary_op}), for an unary operators \texttt{op}, if the operand \texttt{x} is a tainted variable, the result \texttt{y} need be marked as tainted.

In the formula (\ref{formula:binary_op}), \texttt{op} is a binary operator. If one of the two operands \texttt{x} is a tainted variable, the result of their operation \texttt{z} need be marked as a tainted variable.

The Get Element Pointer (GEP) instruction provides a way to calculate pointer offsets. \texttt{base} is the base address to start from and \texttt{addr} is the calculated offset pointer. In the formula (\ref{formula:gep_op}), if \texttt{base} is a tainted pointer, then the derived \texttt{addr} needs to be tainted.

LLVM bitcast instruction converts a value to another type without changing any bits. In the formula (\ref{formula:bitcast_op}), if the original value \texttt{x} is a tainted variable,the bitcast result \texttt{y} should be marked as a tainted variable.


The pseudocode for \texttt{PtrTaint} is Algorithm \ref{algo:ptrTaint}. The value-flow graph (\texttt{vfg}) and the pointer (\texttt{ptr}) that we want to track serve as the input. The output is a collection of tainted sinks. The primary purpose of the algorithm is a hierarchical traversal on the def-use chains corresponding to ptr using a queue. New pointers are continually added to tainted during the traversal process using the aforementioned rules. In line 19, it should be noted that if STELLA discovers a store node and the address is tainted but the address is not, the store node is deemed to be a taint sink and is merged into \texttt{sinks}. \texttt{memcpy} and the LLVM store instruction are both represented as store nodes in VFG.

\begin{algorithm}[t]
\scriptsize
	\caption{Tainting pointers by the propagation rules.}
	\label{algo:ptrTaint}
	\LinesNumbered
	\SetKwFunction{FindSinks}{FindSinks}
	\SetKwFunction{PtrTaint}{PtrTaint}
	\SetKwProg{Fn}{Function}{}{end}	
    \SetKwData{ptr}{ptr}
    \SetKwData{vfg}{vfg}
    \SetKwData{sinks}{sinks}
    \SetKwData{taintedPtrSet}{taintedPtrSet}
    \SetKwComment{Comment}{/* }{ */}
    
    \BlankLine
	\BlankLine
    \KwIn{\ptr: a pointer, \vfg: value flow graph}
    \KwResult{\sinks: taint sinks result}
    \BlankLine
    \BlankLine
    \Fn{\PtrTaint{\ptr,\vfg}}{
        $defNode \gets \texttt{GetDefNode}(\ptr,\vfg)$\\
        $tainted \gets \{\ptr\}$\\
        $visited \gets \{defNode\}$\\
        $sinks \gets \{\}$\\
        $queue.\texttt{Push}(defNode)$\\
        \While{$queue \neq \emptyset$}{
            $curNode \gets queue.\texttt{pop}()$\\
            \For{$node \in \texttt{GetChildNodes}(curNode, vfg)$}{
                $srcPtr \gets \texttt{GetSrcPtr}(node)$\\
                $dstPtr \gets \texttt{GetDstPtr}(node)$\\
                \If{$\texttt{type}(node) \in \{Load,Bitcast,Gep,UnaryOp\}$}{
                        \If{$srcPtr \in tainted$}{
                            $tainted \gets tainted \cup \{dstPtr\}$
                        }
                    }
                    \ElseIf{$\texttt{type}(node) \in \{Store\}$}{
                        \If{$srcPtr \in tainted$}{
                            $tainted \gets tainted \cup \{dstPtr\}$\\
                        }
                        \ElseIf{$srcPtr \not \in tainted $ {\bf and} $ dstPtr \in tainted$}{
                            $sinks \gets sinks \cup \{node\}$\\
                        }
                    }
                    \ElseIf{$\texttt{type}(node) \in \{BinaryOp,PHI\}$}{
                        \If{$node.\texttt{GetSrcPtrs}() \cap tainted \neq \emptyset$}{
                            $tainted \gets tainted \cup \{dstPtr\}$\\
                        }
                    }
                    \If{$node \not \in visited$}{
                        $visited \gets visited \cup \{node\}$\\
                        $queue.\texttt{Push}(node)$\\
                    }
            }
        }
        \KwRet{$sinks$}
    }

\end{algorithm}












\subsubsection{Backward tracking and sensitive information determination.}
After taint sink finding, we can obtain all code snippets in the enclave that transmit information to the untrusted world. Next, we backward track leaked variables and determine whether they are sensitive. Algorithm \ref{algo:backwardTracking} shows our backward tracking algorithm. We process each taint sink, traverse from the VFG node of the sink in the reverse direction along the def-use chain, and get all the paths from the sink to the leaf nodes. Then, we start to find out whether the nodes in the paths are variable allocation instructions (\textit{e.g.} alloc). If so, and they are not yet in the \texttt{InsensitiveTable}, then we add the source code location of the sink and leaked variable to the report.

\begin{algorithm}[t]
\scriptsize
	\caption{Backward tracking}
	\label{algo:backwardTracking}
	\LinesNumbered
	\SetKwFunction{BackwardTracking}{BackTrack}
	\SetKwFunction{AnalyzeOneSink}{BackTrackEachSink}
	\SetKwProg{Fn}{Function}{}{end}	
    \SetKwData{sinks}{sinks}
    \SetKwData{visited}{visited}
    \SetKwData{path}{path}
    \SetKwData{src}{src}
    \SetKwData{vfg}{vfg}
    \SetKwData{report}{report}
    \SetKwComment{Comment}{/* }{ */}
    
    \KwIn{\sinks: taint sinks, \vfg: value flow graph}
    \KwResult{\report: privacy leakage report}
    \Fn{\BackwardTracking{\sinks, \vfg}}{
        \For{$sink \in \sinks$}{
            /*a queue that stores nodes from the sink to leaked variable.*/\\
            $path \gets \texttt{queue}()$\\
            /*a list that stores visited nodes*/\\
            $visited \gets \{\}$\\
            /*backtracking for each sink.*/\\
            $\AnalyzeOneSink(sink, visited, path, vfg)$\\
        }
    }
    \KwIn{\src: source node, \visited: visited nodes, \path: trace path , \vfg: value flow graph}
    \Fn{\AnalyzeOneSink{\src, \visited, \path, \vfg}}{
        /*find a leaked sensitive variable allocation node*/\\
        \If{ $\texttt{type}(\src) \in allocStmts ${ \bf and }$ \src \not \in InsensitiveDataTable$}{
            /*report a leak*/\\
            $\texttt{PrintLeakPath}(src, \path)$\\
            \KwRet{}
        }
        /* avoid encryption or seal functions in order to reduce false positives*/\\
        \If{$\texttt{GetFunc}(src) \in \{encrypt,seal,...\}$}{
            \KwRet{}
        }
        /* push the current node to path and set it visited */\\
        $path.\texttt{Push}(src)$\\
        $visited \gets visited \cup \{src\}$\\
        /*continue visit its parents*/\\
        \For{$parentNode \in \texttt{GetParentNodes}(node, vfg)$}{
            /* If parentNode is not visited, continue to trace up recursively.*/\\
            \If{$parentNode \not \in \visited$}{
                $\AnalyzeOneSink(parentNode,\visited,\path)$\\
            }
        }
        /*reset the node to unvisited for find all leak paths*/\\
        $path.\texttt{Pop}()$\\
        $visited.\texttt{Delete}(src)$\\
    }

\end{algorithm}

\begin{table*}[]
\centering
\caption{The details of selected projects and vulnerabilities we found.}
\small

\begin{tabular}{llcccccc} 
\toprule
\textbf{Project Name}   & \textbf{Description}            & \begin{tabular}[c]{@{}c@{}}\textbf{ECALL }\\\textbf{ user\_check}\end{tabular} & \begin{tabular}[c]{@{}c@{}}\textbf{ECALL}\\\textbf{ out}\end{tabular} & \begin{tabular}[c]{@{}c@{}}\textbf{OCALL}\\\textbf{ in}\end{tabular} & \begin{tabular}[c]{@{}c@{}}\textbf{OCALL}\\\textbf{ return}\end{tabular} & \begin{tabular}[c]{@{}c@{}}\textbf{Null ptr}\\\textbf{ deref}\end{tabular} & \textbf{Total}  \\ 
\midrule
SGX-Tor~\cite{SGX-Tor}                & secure anonymity network        &                                                                                &                                                                       & 1                                                                    & 1                                                                        & 7                                                                           & 9               \\
sgx\_wechat\_app~\cite{sgx_wechat_app}       & trusted wechat app              &                                                                                &                                                                       & 1                                                                    &                                                                          &                                                                            & 1               \\
Fidelius~\cite{Fidelius}               & protect browser users' secrets  &                                                                                &                                                                       & 1                                                                    &                                                                          & 5                                                                          & 6               \\
sgx-based-mix-networks~\cite{SGX-based-mix-networks} & hidden anonymization            &                                                                                & 1                                                                     &                                                                      &                                                                          &                                                                            & 1               \\
sgx-dnet~\cite{SGX-Darknet}               & machine learning inside enclave &                                                                                &                                                                       & 1                                                                    &                                                                          & 25                                                                         & 26              \\
SGX\_SQLite~\cite{SGX_SQLite}            & secure SQLite database          &                                                                                &                                                                       & 1                                                                    &                                                                          &                                                                            & 1               \\
TaLoS~\cite{TaLoS,TaLoS_github}                  & secure TLS library              & 2                                                                              &                                                                       &                                                                      & 1                                                                        &                                                                            & 3               \\
sgx-aes-gcm~\cite{sgx-aes-gcm}            & SGX AES-GCM usage example       &                                                                                & 1                                                                     & 1                                                                    &                                                                          &                                                                            & 2               \\
password-manager~\cite{password-manager}       & password manager using SGX      &                                                                                &                                                                       & 1                                                                    &                                                                          & 1                                                                          & 2               \\
TACIoT~\cite{TACIoT}                 & IoT data protection             &                                                                                &                                                                       & 1                                                                    &                                                                          & 1                                                                          & 2               \\
BiORAM-SGX~\cite{BiORAM-SGX}             & genome analysis system          &                                                                                &                                                                       & 2                                                                    &                                                                          &                                                                            & 2               \\
PrivacyGuard~\cite{PrivacyGuard}           & data analytic inside enclave    &                                                                                &                                                                       & 2                                                                    &                                                                          & 7                                                                          & 9               \\
Town-Crier~\cite{Town_Crier}             & smart contract using SGX        &                                                                                &                                                                       & 2                                                                    &                                                                          & 12                                                                         & 14              \\ 
\hline
\multicolumn{2}{c}{Total}                                 & 2                                                                              & 2                                                                     & 14                                                                   & 2                                                                        & 58                                                                         & 78              \\
\bottomrule
\end{tabular}
\label{tab:one}
\end{table*}


\section{Evaluation}

We have implemented a prototype tool STELLA. STELLA can detect the patterns that cause privacy leakage for enclave programs. We perform evaluation experiments to evaluate the effectiveness and performance of STELLA. We release our tool as open source, and it is available online \footnote{https://anonymous.4open.science/r/STELLA-7724}.
\subsection{Experimental Setup}
We perform experiments to study two research questions:

\textbf{RQ1:} Can STELLA effectively find privacy leakage bugs in enclave programs?

\textbf{RQ2:} How efficient is STELLA when analyzing real-world projects?

We use STELLA to test the enclave programs developed using the SGX SDK on GitHub. We found bugs in 13 popular projects (the average stars of these projects is 30). Table~\ref{tab:one} shows the details of these projects. The code size of these projects varies largely, ranging from a few hundred lines to 527K lines.

Our experiments are done on a ubuntu-20.04 server with an Intel i7-9700T 4.30GHz 8-core CPU. The version of Intel SGX SDK is 2.15. The reported run time is the average of three measurements

\subsection{Effectiveness}

\subsubsection{Overview}

STELLA found 78 vulnerabilities in these projects that could leak sensitive data. Table~\ref{tab:one} shows the number of vulnerabilities with different patterns. For example, we found that SGX-Tor has two privacy leakage bugs, one of which is to write sensitive data to the OCALL \texttt{in} pointer, and the other is to write sensitive data to the pointer returned by OCALL.
\subsubsection{Patterns}
We find all five patterns in real-world enclave projects. Among all patterns, writing private data to OCALL \texttt{in} pointers and to NULL pointers are the most common patterns. Next, we will present a case in real-world projects for each pattern.

\begin{figure}[htbp]
\centering
\includegraphics[]{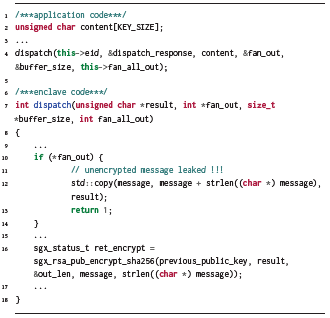}
\caption{A P1 privacy leakage vulnerability in sgx-based-mix-networks. The enclave code mistakenly writes unencrypted sensitive information to the ECALL \texttt{out} pointer (line 12)}
\label{fig:ecall_out}
\end{figure}

\textbf{Write Private Data to ECALL \texttt{out} Pointers (P1)}. Figure \ref{fig:ecall_out} shows a secret disclosure bug in sgx-based-mix-networks. The enclave code incorrectly writes the secret to the ECALL \texttt{out} pointer. From the EDL file, we can know that the parameter \texttt{result} is an \texttt{out} pointer. In \texttt{dispatch} function, \texttt{message} is sensitive data. But at line 12, \texttt{message} is directly copied to \texttt{result} without encryption. When the \texttt{dispatch} function returns, the \texttt{message} plain text in \texttt{result} will be copied to the untrusted \texttt{content} (line 2), causing sensitive information leaked. 

\begin{figure}[htbp]
\centering
\includegraphics[]{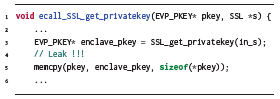}
\caption{A P2 privacy disclosure vulnerability in TaLoS that leaks an SSL private key. \texttt{pkey} is an ECALL \texttt{user\_check} pointer, the enclave code erroneously writes the private key \texttt{enclave\_pkey} to \texttt{pkey}. TaLoS developers have confirmed the vulnerability.}
\label{fig:ecall_user_check}
\end{figure}

\begin{figure}[htbp]
\centering
\includegraphics[]{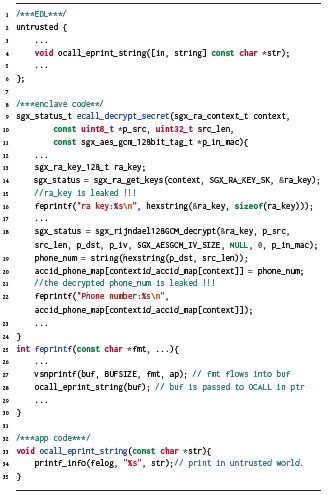}
\caption{The P3 privacy disclosure vulnerability in sgx\_wechat\_app will reveal the sgx\_ra\_key and the decrypted mobile phone number. The enclave code mistakenly passes sensitive information to the OCALL \texttt{in} pointer.}
\label{fig:ocall_in}
\end{figure}

\textbf{Write Private Data to ECALL \texttt{user\_check} pointers (P2)}. TaLoS \cite{TaLoS} is a library that enables applications to terminate TLS connection securely. TaLoS protects sensitive data from disclosure by placing sensitive data within a SGX enclave. But for this library, there is still a hidden danger of SSL private key leakage. Figure \ref{fig:ecall_user_check} demonstrates the code snippet that may cause data leakage. Function \texttt{ecall\_SSL\_get\_privatekey} is an ECALL and the parameter \texttt{pkey} is a \texttt{user\_check} pointer that points outside the enclave. In line 5, \texttt{memcpy} copies the private key from \texttt{enclave\_pkey} to \texttt{pkey}, \textit{i.e.,}the sensitive data is copied from enclave to untrusted world, resulting in privacy leakage. We report this vulnerability and it is included in the CVE-2022-27102.

\textbf{Write Private Data to OCALL \texttt{in} pointers (P3)}. As shown in Figure \ref{fig:ocall_in}, the function \texttt{ecall\_decrypt\_secret} in the enclave converts the variable \texttt{ra\_key} to hexadecimal format and passes it to the function \texttt{feprintf} at line 16. This \texttt{ra\_key} is sensitive data. At line 22, the just decrypted \texttt{phone\_num} is also passed to \texttt{feprintf}. In \texttt{feprintf} function, the \texttt{ra\_key} and \texttt{phone\_num} will be copied into the \texttt{buf} at line 27. Then at line 28, OCALL \texttt{ocall\_eprint\_string} is called by passing \texttt{buf} to its \texttt{in} pointer parameter. During the execution of \texttt{ocall\_eprint\_string}, Intel SGX SDK will copy the sensitive data in \texttt{buf} outside enclave~(line 34). \texttt{ra\_key} and \texttt{phone\_num} are finally printed directly on the standard output of the untrust OS, so attackers can easily get them with little effort.

\begin{figure}[htbp]
\centering
\includegraphics[]{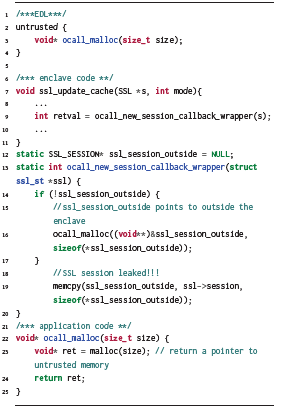}
\caption{A P4 privacy leak in TaLoS. TaLoS applies for a piece of untrusted memory outside the enclave and writes the SSL session to this memory in the enclave. SSL session is important for SSL security, for example, the master key involved in security is included in the SSL session.}
\label{fig:ocall_return}
\end{figure}

\textbf{Write Private Data to OCALL Return Pointers (P4)}. Figure \ref{fig:ocall_return} shows a privacy leakage vulnerability in TaLoS. \texttt{ocall\_malloc} is an OCALL that returns a pointer (line 3). In the enclave code, \texttt{ssl\_update\_cache} updates the cache and calls \texttt{ocall\_new\_sess\\ion\_callback\_wrapper} (line 10). Then, \texttt{ocall\_new\_session\_ca\\llback\_wrapper} will call the OCALL \texttt{ocall\_malloc} when \texttt{ssl\_\\session\_outside} is NULL to apply for a block of untrusted memory (line 17), and write the SSL session to this memory (line 20), resulting in the disclosure of SSL security information.

\begin{figure}[H]
\centering
\includegraphics[]{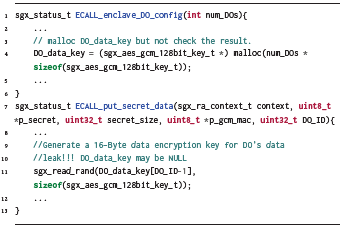}
\caption{A P5 privacy leakage vulnerability in PrivacyGuard. The enclave code uses malloc to allocate a block of memory to store the encryption key, but forgets to check if the return pointer is NULL. when the host runs out of memory or is attacked, the encryption key can be compromised.}
\label{fig:null_ptr_deref}
\end{figure}

\textbf{Write Private Data to NULL Pointers (P5)}. As shown in Figure \ref{fig:null_ptr_deref}, PrivacyGuard's enclave code first uses \texttt{malloc} to apply for a block of trusted memory (line 4) but forgets to check whether the result is successful, so \texttt{DO\_data\_key} may be NULL, and then generates a 16-byte random number as the key. The key is stored in \texttt{DO\_data\_key} (line 11), if the memory allocation fails, the key will be leaked. The developers of PrivacyGuard have confirmed the vulnerability.

\subsection{Performance}

To answer RQ2, we measured the time it took STELLA to analyze these open-source enclaves, and the results are shown in Table \ref{tab:two}. Analysis time is positively related to the number of lines of codes. When the number of lines of code is about 10K, the analysis speed of STELLA is fast, and the analysis can be completed within 10 seconds. When the number of lines of code is around 500K~(TaLoS), the analysis time goes up, but it does not exceed 10 minutes. Overall, STELLA is very efficient even when analyzing large programs.

\begin{table}[htbp]
\caption{The time to analyze each project. The results are average over three measurements.}
\begin{tabular}{@{}lcc@{}}
\toprule
\textbf{Project Name}  & \textbf{Enclave LoC} & \textbf{Time(s)} \\ \midrule
SGX-Tor                & 491,431              & 388.84             \\
sgx\_wechat\_app       & 307                  & 0.23             \\
Fidelius               & 14,129               & 9.11             \\
sgx-based-mix-networks & 211                  & 0.09             \\
sgx-dnet               & 14,344               & 6.86             \\
SGX\_SQLite            & 213,806              & 123.57           \\
TaLoS                  & 527,837              & 407.76           \\
sgx-aes-gcm            & 136                  & 0.01             \\
password-manager       & 6,383                & 0.29             \\
TACIoT                 & 472                  & 0.04             \\
BiORAM-SGX             & 11,251               & 3.59             \\
PrivacyGuard           & 85,015               & 6.33             \\
Town-Crier             & 12,275               & 6.34             \\ \bottomrule
\end{tabular}
\label{tab:two}
\end{table}
\subsection{Threats to Validity}
False positives are not statistically analyzed in the experiment. Many leaks in the report are difficult to distinguish whether it is sensitive information because this requires prior knowledge of the enclave programs. The bugs in Table \ref{tab:one} have been manually confirmed to be high-risk sensitive information. 

In addition, our NULL pointer detection is not sound. we perform pointer analysis~\cite{andersen1994program} (the point-to set of NULL pointer is empty) and \texttt{malloc} analysis (whether the enclave code assumes that \texttt{malloc} will succeed), so the actual numbers of P5 vulnerabilities may be more and the problem is more serious. Furthermore, we cannot detect leaks caused by wild pointers and mathematically manipulated pointers pointing outside the enclave.
\section{Related Work}
Several previous research also studies privacy leakage problems in enclave programs. 

COIN attacks~\cite{khandaker2020coin} summarizes four interface-oriented attacks: Concurrent, Order, Inputs, and Nested, and implements a testing framework to detect bugs with instruction emulation and concolic execution. COIN detects enclave memory information leakage by checking the length of \texttt{memcpy} or a loop condition. This approach can only detect privacy leakage caused by the out-of-bounds copy, but cannot detect the patterns we defined. 

TeeRex~\cite{cloosters2020teerex} mainly detects memory corruption vulnerabilities in the enclave code introduced by the interface between the host and the enclave. These vulnerabilities could allow attackers to corrupt function pointers and arbitrary memory writes. In terms of enclave information leakage, TeeRex's work is relatively limited, and it only briefly explains that under the vulnerability of null pointer dereference, malicious \texttt{user\_check} pointer input causes arbitrary memory read. However, our work demonstrates that even in the absence of malicious third parties, the enclave code may be leaking secrets. 

Moat~\cite{sinha2015moat} employs formal verification to verify whether the enclave code leaks secrets to an adversary. However, Moat is not flexible and scalable enough to apply to large real-world enclave code. STELLA can efficiently analyze large open-source projects.

DEFLECTION~\cite{9505138} verifies the enclave programs by employing compiler instrumentation to insert some privacy security policies into the enclave programs. DEFLECTION introduces runtime overhead for enclave code. In the worst case, the performance overhead is 39.8\%.


\section{Conclusion}
This paper investigates possible privacy leakage in enclave code. The main challenge lies in how to effectively and efficiently identify the privacy leakage code. We at first define five common privacy-leaking patterns, then propose a novel sparse taint analysis method to identify these leaking patterns. We implement a prototype STELLA and analyze several open-source enclave programs on GitHub with STELLA. Our experimental results show that sparse taint analysis can effectively and efficiently detect privacy leaking bugs. We believe that our method will shed light on further research on enclave privacy protection.

\section*{Acknowledgement}
This work was sponsored by CCF- AFSG Research Fund (project id: RF20210017).
\bibliographystyle{ACM-Reference-Format}
\bibliography{stela}
\end{document}